\documentclass{JHEP3}
\usepackage{amsmath}
\usepackage[mathscr]{eucal}
\newcommand{\ket}[1]{|#1\rangle}

\title{SUSY Gauge Theories on Squashed Three-Spheres}

\author{
Naofumi Hama$^a$, Kazuo Hosomichi$^a$ and Sungjay Lee$^b$\\
$^a$Yukawa Institute for Theoretical Physics,\\
~  Kyoto University, Japan
\vskip2mm
$^b$Department of Applied Mathematics and Theoretical Physics,\\
~  University of Cambridge, UK
}

\abstract{We study Euclidean 3D ${\cal N}=2$ supersymmetric gauge
theories on squashed three-spheres preserving isometries $SU(2)\times U(1)$
or $U(1)\times U(1)$. We show that, when a suitable background $U(1)$
gauge field is turned on, these squashed spheres support charged Killing
spinors and therefore ${\cal N}=2$ supersymmetric gauge theories. We
present the Lagrangian and supersymmetry rules for general gauge
theories. The partition functions are computed using localization
principle, and are expressed as integrals over Coulomb branch. For the
squashed sphere with $U(1)\times U(1)$ isometry, its measure and
integrand are identified with the building blocks of structure constants
in Liouville or Toda conformal field theories with $b\ne1$.
}

\preprint{DAMTP-2011-6 \\ YITP-11-1}


\begin{document}

\section{Introduction}\label{sec:intro}

In 3D supersymmetric gauge theories, there has been a remarkable
technical breakthrough based on exact computation of partition functions
of Euclidean theories on three-sphere $S^3$ using localization principle.
It was first developed for superconformal theories by \cite{KWY1, KWY2},
and recently extended to non-conformal theories by \cite{Jafferis, HHL}.
The result is expected to provide us with a new quantitative tool to
study strongly coupled IR physics of 3D supersymmetric theories. In
particular, in \cite{Jafferis} it has  been argued that the 3D partition
function can be used to determine the superconformal R-symmetry at the
infrared fixed points, in a similar spirit to $a$-maximization in 4D
gauge theories. It has also been attracting wide attention from various
other fields, for example Wilson loops \cite{Suy,RS}, 3D dualities
\cite{KWY2}, large-$N$ duality of topological string \cite{MPP} and the
ABJM theory at strong coupling \cite{DT,MP,DMP}. The localization
technique can also be applied to the theory on $S^2\times S^1$ and
allows one to compute the superconformal index. Using this idea, the
superconformal index of ${\cal N}=6$ Chern-Simons matter theories
was computed by \cite{Seok}. It has recently been generalized
to arbitrary ${\cal N}=2$ supersymmetric gauge theories by \cite{IY}.
See also \cite{IYY,CGKP}.

Another aspect, which we will focus on in this paper, is the application
to the AGT relation. In \cite{Pe} the exact partition function
of 4D ${\cal N}=2$ gauge theories on $S^4$ has been worked out,
and it has led to a discovery of a remarkable correspondence
between partition functions of 4D gauge theories and correlation
functions of 2D Liouville or Toda conformal field theories \cite{AGT,Wy}.
In this context, certain 3D supersymmetric gauge theories on $S^3$ play
the role of domain walls \cite{DGG, HLP}, and their partition functions
were shown to correspond to transformation coefficients of conformal
blocks under S-duality (or Moore-Seiberg groupoid operation).

In \cite{HLP} it was observed that the building blocks of 3D partition
functions on round $S^3$, namely the measure over Coulomb branch and the
one-loop determinants, are precisely the building blocks for the structure
constants in Liouville or Toda CFTs with $b=1$. Here $b$ is the coupling
which determines the central charge of Liouville or Toda theories, for
example for Liouville theory one has $c=1+6(b+b^{-1})^2$. In particular,
the one-loop determinant of a charged matter multiplet is given by a
double sine function $s_b(x)$ at $b=1$,
\begin{equation}
 s_b(x)~\equiv~\prod_{m,n\in\mathbb Z_{\ge0}}
 \frac{mb+nb^{-1}+\frac Q2-ix}{mb+nb^{-1}+\frac Q2+ix}\,,
\quad \big(Q\equiv b+\tfrac1b\big)
\end{equation}
and $x$ is related to a coordinate on the Coulomb branch. See
\cite{KLS,BT} for more details on this function. This is plausible if we
recall that 4D ${\cal N}=2$ gauge theories on round $S^4$ are in
correspondence with Liouville or Toda theories with $b=1$.
It is not yet known how to obtain theories with more general $b$ though,
as suggested in \cite{HLP}, a reasonable guess would be that the
background sphere should admit a continuous deformation to account for
this. In this paper we present the answer to this question for 3D theories.

\vskip2mm

We begin by presenting two kinds of squashed $S^3$ in Section
\ref{sec:three-sphere}. The first one appears frequently in the
literature; the metric is written in terms of left-invariant
one-forms and preserves $SU(2)\times U(1)$ symmetry.
The second one is less familiar and preserves only $U(1)\times U(1)$
symmetry, but it has a simple definition as a hyper-ellipsoid embedded
in flat $\mathbb R^4$. For both squashings we show that, if a suitable
background $U(1)$ gauge field is turned on, one can find a pair of
Killing spinors with the $U(1)$ charges $\pm1$ which is necessary for
defining ${\cal N}=2$ supersymmetric gauge theories. We give the general
construction of supersymmetric gauge theories in Section \ref{sec:SUSY}.

We then turn to the computation of partition functions. Section
\ref{sec:partition1} discusses the case of $SU(2)\times U(1)$ symmetric
squashing, where we use the spherical harmonics to work out all the
eigenmodes of the relevant Laplace and Dirac operators. We then use them
to compute the one-loop determinant at each saddle point. Disappointingly,
the partition function turns out to be essentially the same as that for
round sphere, essentially due to $SU(2)$ isometry. 
Next we discuss the less symmetric squashing in Section
\ref{sec:partition2}. Rather than working out the full spectrum of the
Laplace and Dirac operators, we look closely into the structures in
which the bosonic and fermionic modes are paired, and exhaust the modes
unpaired which give nontrivial contributions to the one-loop
determinant. The final expression for the integration measure and
determinant are found to be precisely the building blocks for structure
constants in Liouville or Toda theories with general $b$.

\vskip2mm

We end the introduction by summarizing our conventions for bilinear
products of spinors.
\begin{equation}
 \bar\epsilon\lambda~=~ \bar\epsilon^\alpha C_{\alpha\beta}\lambda^\beta,\quad
 \bar\epsilon\gamma^a\lambda~=~
 \bar\epsilon^\alpha(C\gamma^a)_{\alpha\beta}\lambda^\beta,\quad
 \text{etc}.
\end{equation}
Here $C$ is the charge conjugation matrix with nonzero elements
$C_{12}=-C_{21}=1$, and $\gamma^a$ are Pauli's matrices.
Noticing that $C$ is antisymmetric and $C\gamma^a$ are symmetric,
one finds
\begin{equation}
 \bar\epsilon\lambda=\lambda\bar\epsilon,\quad
 \bar\epsilon\gamma^a\lambda=-\lambda\gamma^a\bar\epsilon
\end{equation}
for all spinors $\bar\epsilon,\lambda$ which we assume to be Grassmann
odd.

\section{Three-Spheres, Round and Squashed}\label{sec:three-sphere}

Here we summarize our notations for various geometric quantities on
round or squashed $S^3$. By squashing we mean certain deformations to
its round metric. We will restrict to those preserving at least
$U(1)\times U(1)$ isometry. The reason is that, when our $S^3$ is
embedded as a domain wall in a 4D space on which an ${\cal N}=2$
supersymmetric gauge theory is defined, the $U(1)\times U(1)$ isometry
is necessary for the omega deformation. There are many metrics
preserving this isometry, of which we take only two simple examples.
After presenting the metrics for these two squashings we show that, with
a suitable background $U(1)$ gauge field turned on, the two squashed
spheres both admit a pair of charged Killing spinors.

The three-sphere is parametrized by an element $g$ of the Lie group
$SU(2)$, and two copies of $SU(2)$ symmetry act on $g$ from the left
and the right. We introduce the left-invariant (LI) and right-invariant
(RI) one-forms
$\mu^a=\mu^a_\nu d\xi^\nu$ and $\tilde\mu^a=\tilde\mu^a_\nu d\xi^\nu$,
\begin{equation}
 g^{-1}dg ~=~ i\mu^a\gamma^a,\quad
 dg g^{-1}~=~ i\tilde\mu^a\gamma^a,
\end{equation}
where $\gamma^a$ are Pauli matrices.
These one-forms satisfy
\begin{equation}
 d\mu^a=\epsilon^{abc}\mu^b\mu^c,\quad
 d\tilde\mu^a=-\epsilon^{abc}\tilde\mu^b\tilde\mu^c.
\end{equation}

\paragraph{Round sphere.}

The left-right invariant metric on the round sphere with radius $\ell$ is
\begin{equation}
 ds^2 ~=~ \tfrac12\ell^2\text{tr}(dgdg^{-1})
 ~=~ \ell^2\mu^a\mu^a~=~\ell^2\tilde\mu^a\tilde\mu^a.
\end{equation}
We define the vielbein in the ``LI frame'' as
$e^a = e^a_\mu d\xi^\mu = \ell\mu^a$.
The spin connection in this frame is
$\omega^{ab}= \varepsilon^{abc}\mu^c$ and satisfies the torsion-free
condition $de^a+\omega^{ab}e^b=0$.
If we define the vielbein from $\tilde\mu^a$ (``RI frame''),
the spin connection is $\tilde\omega^{ab}=-\varepsilon^{abc}\tilde\mu^c$.

Killing spinor $\epsilon$ satisfies the following equation
\begin{equation}
 D\epsilon
 ~\equiv~ d\epsilon+\tfrac14\gamma^{ab}\omega^{ab}\epsilon
 ~=~ e^a\gamma^a\tilde\epsilon,
\end{equation}
for a certain $\tilde\epsilon$.
Here we used $\gamma^{ab}=i\varepsilon^{abc}\gamma^c$.
There are two types of Killing spinors. The first one is constant in the
LI frame,
\begin{equation}
 \epsilon=\epsilon_0~(\text{constant}), \quad
 \tilde\epsilon=+\tfrac i{2\ell}\epsilon.
\end{equation}
The second one reads
\begin{equation}
 \epsilon=g^{-1}\epsilon_0, \quad
 \tilde\epsilon=-\tfrac i{2\ell}\epsilon,
\end{equation}
and is constant in the RI frame.

Let us next introduce the Killing vector fields
$\mathscr{L}^a=\mathscr{L}^{a\mu}\frac\partial{\partial\xi^\mu}$ and
$\mathscr{R}^a=\mathscr{R}^{a\mu}\frac\partial{\partial\xi^\mu}$
which generate the left and the right actions of $SU(2)$.
They can be determined from
\begin{equation}
 \mathscr{L}^ag= i\gamma^ag,\quad
 \mathscr{R}^ag= ig\gamma^a.
\end{equation}
The vector fields $\tfrac i2\mathscr{L}^a$ and $-\tfrac i2\mathscr{R}^a$
satisfy the standard commutation relations of $SU(2)$ Lie algebra.
It is also easy to find
$\mathscr{R}^{a\nu}\mu_\nu^b=\mathscr{L}^{a\nu}\tilde\mu_\nu^b=\delta^{ab}$,
in other words $\mathscr{R}^{a\nu}$ and $\mathscr{L}^{a\nu}$ are proportional
to the inverse-vielbeins in LI or RI frames.
The action of these Killing vector fields on the LI and RI one-forms
is given by
\begin{equation}
 \mathscr{L}^a\tilde\mu^b=2\varepsilon^{abc}\tilde\mu^c,\quad
 \mathscr{R}^a\mu^b=-2\varepsilon^{abc}\mu^c,\quad
 \mathscr{L}^a\mu^b=\mathscr{R}^a\tilde\mu^b=0.
\end{equation}
It therefore follows that $\mu^1\mu^2\mu^3= d^3\xi\text{det}(\mu^a_\nu)$
can be used to define the invariant volume form.

\paragraph{Squashed sphere (familiar one).}

Squashed spheres are defined by the metric or vielbein one-forms
\begin{eqnarray}
&& ds^2 ~=~ \ell^2(\mu^1\mu^1+\mu^2\mu^2)+\tilde\ell^2\mu^3\mu^3,
\nonumber \\
&& (e^1,e^2,e^3)~=~(\ell\mu^1,\ell\mu^2,\tilde\ell\mu^3).
\label{fssgmn}
\end{eqnarray}
The squashed metric preserves the
$SU(2)_{\mathscr{L}}\times U(1)_{\mathscr R^3}$ symmetry.
The torsion-free condition determines the spin connection $\omega^{ab}$
as follows,
\begin{eqnarray}
\omega^{12}&=& (2\tilde\ell^{-1}-f^{-1})e^3,\nonumber \\
\omega^{23}&=& f^{-1}e^1,\nonumber \\
\omega^{31}&=& f^{-1}e^2,
\end{eqnarray}
where $f\equiv \ell^2\tilde\ell^{-1}$ is a constant. Then one can show
that any constant spinor $\psi$ satisfies
\begin{equation}
 d\psi+
 \frac14\gamma^{ab}\omega^{ab}\psi
 ~=~\frac i{2f}\gamma^ae^a\psi
 +i\gamma^3 V\psi,\quad
 V~\equiv~e^3\Big(\frac1{\tilde\ell}-\frac1f\Big).
\label{nks1}
\end{equation}
This is the Killing spinor equation were it not for the last term
in the right hand side.

Now, let us think of the one-form $V$ as a background gauge field
for a certain $U(1)$ symmetry. Let $\epsilon$ be a complex spinor
field with charge $+1$ under this $U(1)$, so that its covariant
derivative becomes
\begin{equation}
 D\epsilon~\equiv~ d\epsilon+\frac14\gamma^{ab}\omega^{ab}\epsilon
 -iV\epsilon.
\end{equation}
Then the constant spinor $\epsilon=(1,0)^t$ satisfies the
($U(1)$-covariant version of) Killing spinor equation,
\begin{equation}
 D\epsilon~=~ \frac{i}{2f}\gamma^ae^a\epsilon.
\end{equation}
Likewise, a spinor $\bar\epsilon$ carrying the $U(1)$ charge $-1$ is
a Killing spinor if $\bar\epsilon=(0,1)^t$.
One also finds
\begin{equation}
\bar\epsilon\epsilon=-1,\quad
\bar\epsilon\gamma^a\epsilon=-\delta^{a3}.
\end{equation}
The latter represents nothing but the Killing vector $\mathcal{R}^3$.
The squashed $S^3$ can be regarded as an $S^1$ fibration over $S^2$,
where each fiber is an orbit of the action of $\mathcal{R}^3$. 

\paragraph{Squashed sphere (less familiar one).}

Next we consider the less familiar version of squashed sphere which
preserve only $U(1)_{\mathscr{L}^3}\times U(1)_{\mathscr{R}^3}$ symmetry.
We take one of the simplest metrics with this property,
\begin{eqnarray}
 ds^2 &=& \ell^2(dx_0^2+dx_1^2)+ \tilde\ell^2(dx_2^2+dx_3^2)\,.
 \nonumber \\&&
 (x_0^2+x_1^2+x_2^2+x_3^2=1)
\end{eqnarray}
Inserting $(x_0,x_1,x_2,x_3)=
(\cos\theta\cos\varphi,\cos\theta\sin\varphi,
 \sin\theta\cos\chi,\sin\theta\sin\chi)$
one obtains
\begin{eqnarray}
ds^2 &=&  f(\theta)^2d\theta^2
 +\ell^2\cos^2\theta d\varphi^2+\tilde\ell^2\sin^2\theta d\chi^2,
 \nonumber \\
 && f(\theta) ~\equiv~
 \sqrt{\ell^2\sin^2\theta+\tilde\ell^2\cos^2\theta}.
\end{eqnarray}
We set the vielbein as follows,
\begin{equation}
 e^1=\ell\cos\theta d\varphi,\quad
 e^2=\tilde\ell\sin\theta d\chi,\quad
 e^3=f(\theta)d\theta.
\label{ss2vb}
\end{equation}
The spin connection then becomes
\begin{equation}
 \omega^{12}=0, \quad
 \omega^{13}=-\frac\ell f\sin\theta d\varphi,\quad
 \omega^{23}= \frac{\tilde\ell}f\cos\theta d\chi.
\label{ss2sc}
\end{equation}

The Killing spinor equation for a spinor field $\psi$ can be reduced to
\begin{eqnarray}
 if\partial_\varphi\psi &=&
 \ell\gamma^2
 (\tfrac12\sin\theta+\cos\theta\partial_\theta)\psi,\nonumber \\
 if\partial_\chi\psi &=&
 \tilde\ell\gamma^1
 (\tfrac12\cos\theta-\sin\theta\partial_\theta)\psi.
\end{eqnarray}
For round sphere one has $\ell=\tilde\ell=f$, and the above equation is
easily shown to have four independent solutions,
\begin{equation}
 \psi_{st}=\left( \begin{array}{r}
 e^{\frac i2(s\chi+t\varphi-st\theta)}\\
-s e^{\frac i2(s\chi+t\varphi+st\theta)}
 \end{array} \right),\qquad
 (s,t=\pm)
\end{equation}
satisfying
\begin{equation}
D_\mu\psi_{st}~=~ -\frac{ist}{2\ell}\gamma_\mu\psi_{st}.
\end{equation}
For squashed spheres with $\ell\neq\tilde\ell$, the same spinor fields
$\psi_{st}$ fail to satisfy the Killing spinor equations.
\begin{equation}
 -\frac{ist}{2f}\gamma_\mu\psi_{st}
 ~=~ D_\mu\psi_{st}-iV_\mu^{(st)}\psi_{st},
\label{ss2ks}
\end{equation}
where
\begin{equation}
 V^{(st)}~=~ \frac t2\Big(1-\frac\ell f\Big)d\varphi
     +\frac s2\Big(1-\frac{\tilde\ell}f\Big)d\chi.
\label{ss2v}
\end{equation}
Again, one can reinterpret the unwanted term in the right hand side
of (\ref{ss2ks}) as the coupling to a background $U(1)$ gauge field.
For the discussions in later sections, we choose to turn on the gauge
field $V=V^{(+-)}$, so that the spinors $\epsilon=\psi_{+-}$ and
$\bar\epsilon=\psi_{-+}$ satisfy the Killing spinor equation with $U(1)$
chatges $\pm1$.

\section{SUSY Theories on Squashed $\bf S^3$}\label{sec:SUSY}

Here we present the Euclidean 3D ${\cal N}=2$ gauge theories on manifolds
with generalized Killing spinors. The formulae for the action and
supersymmery rules are almost the same as those for round $S^3$ (see
e.g. \cite{HHL}), but now contain the background $U(1)$ gauge field we
have turned on. Since the $U(1)$ symmetry rotates the Killing spinors,
it is the R-symmetry which has been gauged. In the following we show
that the closure of supersymmetry algebra and the invariance of
Lagrangian are achieved if the background gauge field is coupled to the
fields according to their R-charge.

\paragraph{Vector multiplets.}

Vector multiplet fields obey the following transformation laws,
\begin{eqnarray}
 \delta A_\mu &=&
 -\tfrac i2(\bar\epsilon\gamma_\mu\lambda-\bar\lambda\gamma_\mu\epsilon),
 \nonumber \\
 \delta\sigma &=&
 \tfrac12(\bar\epsilon\lambda-\bar\lambda\epsilon),
 \nonumber \\
 \delta\lambda &=&
 \tfrac12\gamma^{\mu\nu}\epsilon F_{\mu\nu}-D\epsilon
      +i\gamma^\mu\epsilon D_\mu\sigma
 +\tfrac{2i}3\sigma\gamma^\mu D_\mu\epsilon,
 \nonumber \\
 \delta\bar\lambda &=&
 \tfrac12\gamma^{\mu\nu}\bar\epsilon F_{\mu\nu}+D\bar\epsilon
                 -i\gamma^\mu\bar\epsilon D_\mu\sigma
 -\tfrac{2i}3\sigma\gamma^\mu D_\mu\bar\epsilon,
 \nonumber \\
 \delta D &=&
 -\tfrac i2\bar\epsilon\gamma^\mu D_\mu\lambda
 -\tfrac i2D_\mu\bar\lambda\gamma^\mu\epsilon
 +\tfrac i2[\bar\epsilon\lambda,\sigma]
 +\tfrac i2[\bar\lambda\epsilon,\sigma]
 -\tfrac i6(D_\mu\bar\epsilon\gamma^\mu\lambda
         +\bar\lambda\gamma^\mu D_\mu\epsilon).
\label{trvec}
\end{eqnarray}
Here and throughout this paper, $D_\mu$ denotes the covariant derivative
with respect to gauge, local Lorentz and background gauged R-symmetries
as well as general covariance. $\gamma^\mu=e^\mu_a\gamma^a$ is the Dirac
matrix with curved index.
The spinors $\epsilon,\bar\epsilon$ are assumed to satisfy
Killing spinor equation. Namely there are spinors
$\tilde\epsilon,\tilde{\bar\epsilon}$ which satisfy
\begin{eqnarray}
 D_\mu\epsilon &\equiv&
 (\partial_\mu+\tfrac14\omega_\mu^{ab}\gamma^{ab}-iV_\mu)\epsilon
 ~=~ \gamma_\mu\tilde\epsilon,\nonumber \\
 D_\mu\bar\epsilon &\equiv&
 (\partial_\mu+\tfrac14\omega_\mu^{ab}\gamma^{ab}+iV_\mu)\bar\epsilon
 ~=~ \gamma_\mu\tilde{\bar\epsilon}.
\label{KSeq1}
\end{eqnarray}
Denoting $\delta$ as the sum of unbarred and barred parts,
$\delta=\delta_\epsilon+\delta_{\bar\epsilon}$, one can show that
two unbarred or two unbarred supersymmetries commute. Also, on most of
the fields the commutator $[\delta_\epsilon,\delta_{\bar\epsilon}]$ is a
sum of translation by $iv^\mu$, gauge transformation by $\Lambda$,
Lorentz rotation by $\Theta^{\mu\nu}$, dilation by $\rho$ and R-rotation
by $\alpha$, where
\begin{eqnarray}
 v^\mu &=& \bar\epsilon\gamma^\mu\epsilon,
 \nonumber \\
 \Theta^{\mu\nu} &=& iD^{[\mu}v^{\nu]}+iv^\lambda\omega_\lambda^{\mu\nu},
 \nonumber \\
 \Lambda &=& -iA_\mu v^\mu+\sigma\bar\epsilon\epsilon,
 \nonumber \\
 \rho &=& \tfrac i3
 (\bar\epsilon\gamma^\mu D_\mu\epsilon
 +D_\mu\bar\epsilon\gamma^\mu\epsilon),
 \nonumber \\
 \alpha &=& \tfrac
 i3(D_\mu\bar\epsilon\gamma^\mu\epsilon-\bar\epsilon\gamma^\mu D_\mu\epsilon)
 +v^\mu V_\mu.
\label{sypar}
\end{eqnarray}
The only exception is that
\begin{eqnarray}
 [\delta_\epsilon,\delta_{\bar\epsilon}]D &=&
 iv^\mu\partial_\mu D + i[\Lambda,D]+2\rho D
 \nonumber \\&&
 +\tfrac13\sigma
 (\bar\epsilon\gamma^\mu\gamma^\nu D_\mu D_\nu\epsilon
 -\epsilon\gamma^\mu\gamma^\nu D_\mu D_\nu\bar\epsilon).
\end{eqnarray}
The last term in the right hand side vanishes provided that
$\epsilon$ and $\bar\epsilon$ satisfy, in addition to (\ref{KSeq1}), the
following equations
\begin{eqnarray}
 \gamma^\mu\gamma^\nu D_\mu D_\nu\epsilon &=&
 -\tfrac38(R+2iV_{\mu\nu}\gamma^{\mu\nu})\epsilon,
 \nonumber \\
 \gamma^\mu\gamma^\nu D_\mu D_\nu\bar\epsilon &=&
 -\tfrac38(R-2iV_{\mu\nu}\gamma^{\mu\nu})\bar\epsilon
\label{kseq2}
\end{eqnarray}
for a certain set of functions $(R,V_{\mu\nu})$. By arguing in a similar
way to \cite{HHL} one finds that $R$ is the scalar curvature of the 3D
manifold and $V_{\mu\nu}\equiv\partial_\mu V_\nu-\partial_\nu V_\mu$ is
the field strength of the background gauge field. Note that the Killing
spinors on the squashed $S^3$ of our interest actually all satisfy a
stronger condition (\ref{nscd}), so the supersymmetry is not reduced
by the above conition (\ref{kseq2}). For later convenience,
we give here some additional formulae which are related to (\ref{kseq2}).
\begin{eqnarray}
 \gamma^\mu\gamma^\nu D_\mu D_\nu\epsilon &=& 3D_\mu D^\mu\epsilon,\nonumber\\
 \gamma^\mu\gamma^\nu D_\mu D_\nu\bar\epsilon &=& 3D_\mu D^\mu\bar\epsilon,
\end{eqnarray}

\paragraph{Matter multiplets.}

The fields in a chiral multiplet coupled to a gauge symmetry
transform as follows,
\begin{eqnarray}
 \delta\phi &=& \bar\epsilon\psi, \nonumber \\
 \delta\bar\phi &=& \epsilon\bar\psi, \nonumber \\
 \delta\psi &=& i\gamma^\mu\epsilon D_\mu\phi +i\epsilon\sigma\phi
 +\tfrac{2qi}3\gamma^\mu D_\mu\epsilon\phi+\bar\epsilon F,
 \nonumber \\
 \delta\bar\psi &=& i\gamma^\mu\bar\epsilon D_\mu\bar\phi
 +i\bar\phi\sigma\bar\epsilon+\tfrac{2qi}3\bar\phi\gamma^\mu D_\mu\bar\epsilon
 +\bar F\epsilon,
 \nonumber \\
 \delta F &=&
 \epsilon(i\gamma^\mu D_\mu\psi-i\sigma\psi-i\lambda\phi)
 +\tfrac i3(2q-1)D_\mu\epsilon\gamma^\mu\psi,
 \nonumber \\
 \delta\bar F &=&
 \bar\epsilon(i\gamma^\mu D_\mu\bar\psi-i\bar\psi\sigma+i\bar\phi\bar\lambda)
 +\tfrac i3(2q-1)D_\mu\bar\epsilon\gamma^\mu\bar\psi.
\label{dncm}
\end{eqnarray}
Here we assumed the fields $\phi,\psi,F$ ($\bar\phi,\bar\psi,\bar F$) to be
column vectors (resp. row vectors) on which the vector multiplet fields
act as matrices from the left (right).
The lowest components $(\phi,\bar\phi)$ are assigned the dimension $q$
and R-charge $(-q,+q)$, as one can obtain from the supersymmetry
algebra realized on these fields.
The supersymmetry algebra can be easily shown to close off-shell,
except that two unbarred supersymmetries do not simply commute on $F$,
\begin{equation}
 [\delta_\epsilon,\delta_{\epsilon'}]F ~=~
 \epsilon\gamma^{\mu\nu}\epsilon'(2D_\mu D_\nu\phi+iF_{\mu\nu}\phi)
+\tfrac{2q}3\phi\cdot
 (\epsilon \gamma^\mu\gamma^\nu D_\mu D_\nu\epsilon'
 -\epsilon'\gamma^\mu\gamma^\nu D_\mu D_\nu\epsilon).
\end{equation}
The right hand side does not vanish for general Killing spinors
$\epsilon,\epsilon'$. With an additional condition (\ref{kseq2})
one finds
\begin{equation}
 [\delta_\epsilon,\delta_{\epsilon'}]F ~=~
 \epsilon\gamma^{\mu\nu}\epsilon'
 (2D_\mu D_\nu\phi+iF_{\mu\nu}\phi-iq\phi V_{\mu\nu}).
\end{equation}
The right hand side vanishes if $\phi$ couples to $V_\mu$ according
to its R-charge $-q$, namely,
\begin{equation}
 D_\mu\phi\equiv (\partial_\mu-iA_\mu+iqV_\mu)\phi.
\end{equation}
Likewise, two barred supersymmetry commute on $\bar F$ provided that
the Killing spinors satisfy (\ref{kseq2}) and $\bar\phi$ couples
with $V_\mu$ according to its charge $+q$.

\paragraph{Supersymmetric Lagrangians.}

The Chern-Simons Lagrangian for ${\cal N}=2$ vector multiplet is invariant
under supersymmetry.
\begin{equation}
 {\cal L}_{\rm CS} ~=~ {\rm Tr}\left[
   \tfrac1{\sqrt g}\varepsilon^{\mu\nu\lambda}
  (A_\mu\partial_\nu A_\lambda-\tfrac{2i}3A_\mu A_\nu A_\lambda)
   -\bar\lambda\lambda+2D\sigma  \right].
\end{equation}
The $F$-term of gauge-invariant chiral multiplets of R-charge $q=2$
is invariant under supersymmetry up to total derivatives.
\begin{equation}
 \delta F ~=~ iD_\mu(\epsilon\gamma^\mu\psi),\quad
 \delta\bar F ~=~ iD_\mu(\bar\epsilon\gamma^\mu\bar\psi).
\end{equation}
These terms are invariant under $\delta$ for any Killing spinors
$\epsilon,\bar\epsilon$.
In addition, chiral matter multiplets with canonical dimensions have
the kinetic Lagrangian,
\begin{eqnarray}
 {\cal L} &=& D_\mu\bar\phi D^\mu\phi -i\bar\psi\gamma^\mu D_\mu\psi
 +\tfrac R8\bar\phi\phi+i\bar\psi\sigma\psi
 \nonumber \\ &&
 +i\bar\psi\lambda\phi-i\bar\phi\bar\lambda\psi
 +i\bar\phi D\phi+\bar\phi\sigma^2\phi+\bar FF,
\label{Lcm}
\end{eqnarray}
which is invariant under supersymmetry if the Killing spinors
$\epsilon,\bar\epsilon$ satisfy (\ref{kseq2}).

There are Lagrangians which are not superconformal but are
still invariant under some supersymmetry. In the following we look for
the quantities which are invariant if the parameters
$\epsilon,\bar\epsilon$ satisfy
\begin{equation}
 D_\mu\epsilon=\tfrac{i}{2f}\gamma_\mu\epsilon,\quad
 D_\mu\bar\epsilon=\tfrac{i}{2f}\gamma_\mu\bar\epsilon
\label{nscd}
\end{equation}
for some function $f$.
Note that by combining these with (\ref{kseq2}) one finds
\begin{eqnarray}
 (R+2i\gamma^{\mu\nu}V_{\mu\nu})\epsilon &=&
 (6f^{-2}-4i\gamma^\mu\partial_\mu f^{-1})\epsilon,\nonumber\\
 (R-2i\gamma^{\mu\nu}V_{\mu\nu})\bar\epsilon &=&
 (6f^{-2}-4i\gamma^\mu\partial_\mu f^{-1})\bar\epsilon.
\end{eqnarray}
One example for such Lagrangian is the kinetic Lagrangian
for matter fields with non-canonical R-charges.
\begin{eqnarray}
 {\cal L}_\text{mat} &=&
  D_\mu\bar\phi D^\mu\phi
 +\bar\phi\sigma^2\phi
 +\tfrac{i(2q-1)}f\bar\phi\sigma\phi
 -\tfrac{q(2q-1)}{2f^2}\bar\phi\phi
 +\tfrac q4R\bar\phi\phi
 +i\bar\phi D\phi
 +\bar FF
 \nonumber \\ &&
 -i\bar\psi\gamma^\mu D_\mu\psi
 +i\bar\psi\sigma\psi
 -\tfrac{(2q-1)}{2f}\bar\psi\psi
 +i\bar\psi\lambda\phi
 -i\bar\phi\bar\lambda\psi.
\label{Lncm}
\end{eqnarray}
Another example is the Yang-Mills Lagrangian for vector multiplet.
\begin{eqnarray}
 {\cal L}_\text{YM} &=& \text{Tr}\Big(
  \tfrac14F_{\mu\nu}F^{\mu\nu}+\tfrac12D_\mu\sigma D^\mu\sigma
 +\tfrac12(D+\tfrac\sigma f)^2
 \nonumber \\ && ~~~
 +\tfrac i2\bar\lambda\gamma^\mu D_\mu\lambda
 +\tfrac i2\bar\lambda[\sigma,\lambda]
 -\tfrac1{4f}\bar\lambda\lambda
 \Big).
\end{eqnarray}
Note that ${\cal L}_\text{mat}$ and ${\cal L}_\text{YM}$ can be
expressed as total-superderivatives,
\begin{eqnarray}
 \bar\epsilon\epsilon\cdot{\cal L}_\text{mat} &=&
 \delta_{\bar\epsilon}\delta_\epsilon\Big(
  \bar\psi\psi-2i\bar\phi\sigma\phi+\tfrac{2(q-1)}{f}\bar\phi\phi
 \Big),
 \nonumber \\
 \bar\epsilon\epsilon\cdot{\cal L}_\text{YM} &=&
 \delta_{\bar\epsilon}\delta_\epsilon\text{Tr}\Big(
  \tfrac12\bar\lambda\lambda-2D\sigma \Big).
\label{Ltot}
\end{eqnarray}
Finally, there is an analogue of FI D-term for abelian vector multiplet.
The auxiliary field $D$ in abelian vector multiplet is also invariant
up to total derivative,
\begin{equation}
 {\cal L}_\text{FI}\equiv D-\tfrac\sigma f,\quad
 \delta {\cal L}_\text{FI} =
 -\tfrac i2D_\mu(\bar\epsilon\gamma^\mu\lambda
                       +\bar\lambda\gamma^\mu\epsilon).
\end{equation}

\section{Partition function on Squashed $\bf S^3$ (Familiar One)}
\label{sec:partition1}

Here we compute partition functions of supersymmetric gauge theories
on squashed $S^3$ based on localization principle. As has been explained
in \cite{KWY1,KWY2}, the path integral localizes onto the saddle points
where the supersymmetry variation of all the fermions vanish.
They are characterized by
\begin{equation}
 A_\mu = \phi= 0,\quad
 \sigma=-f D= \text{constant}.
\end{equation}
The integration over all the modes transverse to the locus of saddle
points can be made finite by introducing arbitrary weight $e^{-S}$,
where the regulator action $S$ should be supersymmery exact. Our
Lagrangians ${\cal L}_\text{mat}$ and ${\cal L}_\text{YM}$ are both
supersymmetry exact so they can be included in $S$ with arbitrary
coefficient. When this coefficient is taken larger and larger, then the
saddle-point (Gaussian) approximation for the path integral becomes more
and more accurate. The partition function can therefore be computed
by truncating the regulator action up to quadratic order at each saddle
point, evaluating the Gaussian path integral (one-loop determinant) and
then integrating over the space of saddle points parametrized by $\sigma$.

In the following we calculate these determinants for chiral and vector
multiplets on two different versions of squashed $S^3$.
In this section we focus on the familiar squashing preserving
$SU(2)_\mathscr{L}\times U(1)_{\mathscr{R}^3}$ symmetry, for which the
metric is given by (\ref{fssgmn}) and the background $U(1)$ gauge field
$V$ is given in (\ref{nks1}).

\paragraph{Matter multiplets.}

Let us evaluate the determinant from matter chiral multiplets first. For
simplicity, we focus on the simplest example of a single chiral
multiplet of R-charge $q$ coupled to an abelian vector multiplet. The
matter kinetic term on the saddle points is ${\cal L}_\phi+{\cal L}_\psi$, with
\begin{eqnarray}
 {\cal L}_\phi &=&
 g^{\mu\nu}D_\mu\bar\phi D_\nu\phi +\bar\phi\sigma^2\phi
 +\tfrac{2i(q-1)}f\bar\phi\sigma\phi
 +\Big\{\tfrac{q(1-2q)}{2f^2}+\tfrac{qR}{4}\Big\}\bar\phi\phi,
 \nonumber \\
 {\cal L}_\psi &=&
 -i\bar\psi\gamma^\mu D_\mu\psi
 +i\bar\psi\sigma\psi
 -\tfrac{2q-1}{2f}\bar\psi\psi.
\end{eqnarray}
We rewrite them using the differential operators $\mathscr{R}^a$
as well as $R=\frac8{f\tilde\ell}-\frac2{f^2}$ and get
\begin{eqnarray}
 {\cal L}_\phi &=&
 \tfrac1{\ell^2}
 \Big( \mathscr{R}^1\bar\phi\mathscr{R}^1\phi
      +\mathscr{R}^2\bar\phi\mathscr{R}^2\phi \Big)
+\tfrac1{\tilde\ell^2}
 \Big\{\mathscr{R}^3\bar\phi-iq(1-\tfrac{\tilde\ell}f)\bar\phi\Big\}
 \Big\{\mathscr{R}^3\phi+iq(1-\tfrac{\tilde\ell}f)\phi\Big\}
 \nonumber \\&&
 +\bar\phi\Big(\sigma^2+\tfrac{2i(q-1)\sigma}f-\tfrac{q^2}{f^2}
 +\tfrac{2q}{f\tilde\ell}\Big)\phi,
 \nonumber \\
 {\cal L}_\psi &=&
 \bar\psi\left\{
  -\tfrac i\ell(\gamma^1\mathscr R^1+\gamma^2\mathscr R^2)
  -\tfrac i{\tilde\ell}\gamma^3\mathscr R^3
  +\tfrac{1-q}{f}+\tfrac1{\tilde\ell}
  +(q-1)(\tfrac1{\tilde\ell}-\tfrac1f)\gamma^3
  +i\sigma
 \right\}\psi.
\end{eqnarray}
We further rewrite them in terms of $J^a=\frac1{2i}\mathscr{R}^a$ and
$S^a=\frac12\gamma^a$ satisfying standard $SU(2)$ commutation relations,
and obtain the Laplace operator $\Delta_\phi$ for scalar field and
the Dirac operator $\Delta_\psi$ for spinor field,
\begin{eqnarray}
 \Delta_\phi
 &=&
  \frac4{\ell^2}(J^1J^1+J^2J^2)
 +\frac4{\tilde\ell^2}\Big\{J^3+\frac q2(1-\frac{\tilde\ell}f)\Big\}^2
 +\sigma^2+\frac{2i(q-1)\sigma}f-\frac{q^2}{f^2}
 +\frac{2q}{f\tilde\ell},
 \nonumber \\
 \Delta_\psi &=&
 \frac4\ell(S^1J^1+S^2J^2)+\frac4{\tilde\ell}S^3J^3
 +\frac1{\tilde\ell}+\frac{1-q}f+2(q-1)\Big(\frac1{\tilde\ell}-\frac1f\Big)S^3.
\end{eqnarray}
The one-loop determinant can thus be computed from the spectrum of these
operators.

The Laplace operator is diagonalized by scalar spherical harmonics which
belong to the representations $(j,j)$ of
$SU(2)_\mathscr{R}\times SU(2)_\mathscr{L}$. There are therefore $2j+1$
scalar wave functions with angular momentum $j$ and $J^3=m$ corresponding to
the eigenvalue
\begin{eqnarray}
 \Delta_\phi
 &=&
  \frac{4j(j+1)-4m^2}{\ell^2}
 +\frac{\{2m+q(1-\frac{\tilde\ell}f)\}^2}{\tilde\ell^2}
 +\sigma^2+\frac{2i(q-1)\sigma}f-\frac{q^2}{f^2}
 +\frac{2q}{f\tilde\ell},
\label{Dphi1}
\end{eqnarray}
Note that for $m=j$ this can be factorized as follows,
\begin{equation}
\Delta_\phi =
  \left(\frac{2j+q}{\tilde\ell}+\frac{2-2q}{f}+i\sigma\right)
  \left(\frac{2j+q}{\tilde\ell}-i\sigma\right).
\end{equation}

Next we turn to the spectrum of Dirac operator $\Delta_\psi$.
Its generic eigenstates are suitable linear combinations of two states
$\ket{j;m,\frac12}$ and $\ket{j;m+1,-\frac12}$, where $j$ labels the
orbital angular momentum and
\[
 \begin{array}{rcr}
 J^3\ket{j;m,s}&=& m\ket{j;m,s}, \\
 S^3\ket{j;m,s}&=& s\ket{j;m,s},
 \end{array}\quad
 \begin{array}{lcl}
 (J^1\pm iJ^2)\ket{j;m,s}&=& (j\mp m)\ket{j;m\pm1,s}, \\
 (S^1\pm iS^2)\ket{j;m,\mp\tfrac12}&=& \ket{j;m,\pm\tfrac12}.
 \end{array}
\]
The action of $\Delta_\psi$ on the states of the form
\begin{equation}
 x_+\ket{j;m,\tfrac12}+x_-\ket{j,m+1,-\tfrac12},
\end{equation}
can be translated into the following $2\times2$ matrix acting on $(x_+,x_-)^t$.
\[
 \left(\begin{array}{cc}
  \frac{2m+1}{\tilde\ell}+(q-1)(\frac1{\tilde\ell}-\frac1f)
 +\frac{1-q}f+i\sigma &
 \frac2\ell(j+m+1) \\ \frac2\ell(j-m) &
 -\frac{2m+1}{\tilde\ell}-(q-1)(\frac1{\tilde\ell}-\frac1f)
 +\frac{1-q}f+i\sigma
 \end{array}\right).
\]
Its determinant is precisely $(-1)$ times the expression (\ref{Dphi1})
for the eigenvalue of $\Delta_\phi$. Note that $m$ can take values
between $-j$ to $j-1$. In addition to these generic eigenstates, the
states $\ket{j,\pm j,\pm\frac12}$ are themselves the eigenstates of
$\Delta_\psi$ for the eigenvalues
\begin{equation}
 \Delta_\psi~=~
 \frac{2j+q}{\tilde\ell}+\frac{2-2q}f+i\sigma,\quad
 \frac{2j+2-q}{\tilde\ell}+i\sigma.
\end{equation}
Since $SU(2)_\mathscr{L}$ is unbroken, all these spectra aquire
the multiplicity $(2j+1)$.

Combining everything together, we obtain the one-loop determinant,
\begin{eqnarray}
 \frac{\text{det}\Delta_\psi}{\text{det}\Delta_\phi} &=&
 \prod_{2j\in\mathbb{Z}_{\ge0}}
 \left[
 (-1)^{2j}
 \frac{
  (\frac{2j+q}{\tilde\ell}+\frac{2-2q}f+i\sigma)
  (\frac{2j+2-q}{\tilde\ell}+i\sigma)}{
  (\frac{2j+q}{\tilde\ell}+\frac{2-2q}f+i\sigma)
  (\frac{2j+q}{\tilde\ell}-i\sigma) }
 \right]^{2j+1}
 \nonumber \\ &=&
 \prod_{n>0}
 \left(\frac{n+1-q+i\tilde\ell\sigma}{n-1+q-i\tilde\ell\sigma}\right)^n
 ~=~ s_{b=1}(i-iq-\tilde\ell\sigma)\,.
\label{detmat1}
\end{eqnarray}
This is essentially the same as the result for round $S^3$ (see, e.g.
\cite{HHL}) except that the radius of round $S^3$ is replaced by
$\tilde\ell$.

\paragraph{Vector multiplets.}

Next we study vector multiplets. We denote by $\varphi$ the fluctuation
mode of the scalar field away from its classical value $\sigma$, and
consider the path integral with linearized Lagrangian
${\cal L}={\cal L}_B+{\cal L}_F$,
\begin{eqnarray}
 {\cal L}_B &=&
 \text{Tr}\Big(\tfrac14
 \hat F_{\mu\nu}\hat F^{\mu\nu}
 +\tfrac12\partial_\mu\varphi\partial^\mu\varphi
 -\tfrac12[A_\mu,\sigma][A^\mu,\sigma]
 -i[A_\mu,\sigma]\partial^\mu\varphi\Big),
 \nonumber \\
 {\cal L}_F &=&
 \text{Tr}\Big(
  \tfrac i2\bar\lambda\gamma^\mu D_\mu\lambda
 +\tfrac i2\bar\lambda[\sigma,\lambda]
 -\tfrac1{4f}\bar\lambda\lambda  \Big),
\label{Llin}
\end{eqnarray}
where $\hat F_{\mu\nu}\equiv\partial_\mu A_\nu-\partial_\nu A_\mu$.
We decompose all the adjoint-valued fields with respect to the
Cartan-Weyl basis $(H_i,E_\alpha,E_{-\alpha})$ satisfying the
commutation relations,
\[
 [H_i,H_j]=0,\quad
 [H_i,E_\alpha]=\alpha_i E_\alpha,\quad
 [E_\alpha,E_{-\alpha}]=\tfrac2{|\alpha|^2}\alpha_iH_i.
\]
We may assume that $\sigma$ takes values in the Cartan subalgebra, i.e.
$\sigma=\sigma_iH_i$. Then the Lagrangian for fermions can be rewritten
into the form
\begin{equation}
 {\cal L}_F~=~
 \sum_i\bar\lambda_i(i\gamma^\mu D_\mu-\tfrac1{2f})\lambda_i
+\sum_{\alpha\in\Delta}
 \bar\lambda_{-\alpha}(i\gamma^\mu D_\mu+i\sigma\alpha-\tfrac1{2f})
 \lambda_\alpha,
\end{equation}
where $\sigma\alpha\equiv\sigma_i\alpha_i$. Note that $D_\mu\lambda$
contains the couplings to spin connection as well as the background
$U(1)$ vector field according to the R-charge of $\lambda$. The Dirac
operator for these fermions is thus the same as that of matter fermions
in a chiral multiplet with $q=0$. The determinant of Dirac operator for
the fields $\lambda_{\pm\alpha},\bar\lambda_{\pm\alpha}$ is thus given by
\begin{eqnarray}
&&
 \prod_{j\ge0}
\left[
 \Big(\tfrac{2j}{\tilde\ell}+\tfrac2f-i\sigma\alpha\Big)
 \Big(\tfrac{2j}{\tilde\ell}+\tfrac2f+i\sigma\alpha\Big)
 \Big(\tfrac{2j+2}{\tilde\ell}-i\sigma\alpha\Big)
 \Big(\tfrac{2j+2}{\tilde\ell}+i\sigma\alpha\Big)
\right]^{2j+1}
\nonumber \\ && \times
 \prod_{j\ge1/2}
 \prod_{m=-j}^{j-1}
 \left[ \Delta_0(j,m)+(\sigma\alpha)^2+\tfrac{2i}f\sigma\alpha\right]^{2j+1}
 \left[ \Delta_0(j,m)+(\sigma\alpha)^2-\tfrac{2i}f\sigma\alpha\right]^{2j+1},
\end{eqnarray}
where
\begin{equation}
\Delta_0(j,m)\equiv
  \frac{4j(j+1)-4m^2}{\ell^2}
 +\frac{4m^2}{\tilde\ell^2}.
\label{Delta0}
\end{equation}

The bosonic Lagrangian is easier to handle once written in terms
of differential forms,
\begin{equation}
 d^3\xi\sqrt{g}{\cal L}_B~=~
 \tfrac12\text{Tr}\Big(dA\wedge\ast dA+d\varphi\wedge\ast d\varphi
 -[\sigma,A]\wedge[\sigma,\ast A]+i[\sigma,A]\wedge\ast d\varphi\Big).
\end{equation}
This leads to the Laplace operator acting on a pair $(A,\varphi)$
as follows,
\begin{equation}
 \Delta_B~:~\left(\begin{array}c A\\ \varphi\end{array}\right)~\mapsto
 \left(\begin{array}{l}
 \ast d\ast dA+[\sigma,[\sigma,A]]-i[\sigma,d\varphi]\\
 -i[\sigma,\ast d\ast A]-\ast d\ast d\varphi
 \end{array}\right).
\end{equation}
Here the Hodge star $\ast$ is a linear map from $d$-forms
to $(3-d)$-forms satisfying $\ast\ast=1$.
It is defined by the equations
\begin{equation}
\ast 1=e^1e^2e^3,\quad
\ast e^1=e^2e^3,\quad
\ast e^2=e^3e^1,\quad
\ast e^3=e^1e^2.
\end{equation}
Let us hereafter focus on the components of $(A,\varphi)$ proportional to
the Lie algebra element $E_\alpha$, so that the commutator with $\sigma$
becomes just the multiplication by $\sigma\alpha$.

Generic eigenmodes of $\Delta_B$ take the form
\begin{equation}
 A_\alpha = x^+Y_{j,m+1}e^- + x^-Y_{j,m-1}e^+ + x^3Y_{j,m}e^3,\quad
 \varphi_\alpha = x Y_{j,m},
\end{equation}
where $e^\pm\equiv\frac12(e^1\pm ie^2)$, and the spherical
harmonics $Y_{j,m}$ is normalized to satisfy
\[
 J^3Y_{j,m}=mY_{j,m},\quad
 (J^1\pm iJ^2)Y_{j,m}=(j\mp m)Y_{j,m\pm1}.
\]
The Laplace operator $\Delta_B$ then translates into a
matrix $X$ acting on $(x^+,x^3,x^-,x)^t$.
It takes the $(3+1)\times(3+1)$ block decomposed form
\begin{equation}
 X~=~
 \left(\begin{array}{cc}U^2+(\sigma\alpha)^2 &
 ~~(\sigma\alpha)\vec v \\
   (\sigma\alpha)\vec w & \Delta_0
\end{array}\right),
\end{equation}
with
\begin{eqnarray}
 U&=& \left(\begin{array}{ccc}
 \displaystyle -\frac{2m}{\tilde\ell} &
 \displaystyle \frac{2(j-m)}\ell & 0 \\
 \displaystyle \frac{j+m+1}\ell &
 \displaystyle \frac2f &
 \displaystyle -\frac{j-m+1}\ell \\
 0 &
 \displaystyle -\frac{2(j+m)}\ell &
 \displaystyle \frac{2m}{\tilde\ell}
 \end{array}\right),
 \nonumber \\
 \vec v &=&
 \left(\frac{2(j-m)}\ell,\frac {2m}{\tilde\ell},\frac{2(j+m)}\ell\right)^t,
 \nonumber \\
 \vec w &=&
 \left(\frac{j+m+1}\ell,\frac{2m}{\tilde\ell},\frac{j-m+1}\ell\right),
\end{eqnarray}
and $\Delta_0$ given in (\ref{Delta0}).
Noticing that $U\vec v=\vec wU=0$ and $\Delta_0=\vec w\cdot\vec v$,
one can easily find two eigenvectors of $X$,
\begin{eqnarray}
 \big(\vec v,-(\sigma\alpha)\big)^t&~:~&X=0,\nonumber \\
 \big((\sigma\alpha)\vec v,\Delta_0\big)^t&~:~&X=\Delta_0+(\sigma\alpha)^2.
\end{eqnarray}
These two modes are longitudinal in the sense that $A$ is a total derivative.
The above two eigenvalues do not enter the formula for one-loop
determinant, as we explain later. The other two eigenvalues of $X$
correspond to transverse modes, for which $\varphi\equiv0$ and $A$ is
divergenceless. They have eigenvalues  $X=\zeta_1^2+(\sigma\alpha)^2$
and $\zeta_2^2+(\sigma\alpha)^2$, where $\zeta_1,\zeta_2$ are the
two nonzero eigenvalues of $U$. They can therefore be found as two
nonzero roots of the characteristic equation
\begin{equation}
 0~=~ \text{Det}(\zeta-U)~=~\zeta^3-\frac2f\zeta^2-\Delta_0\zeta.
\end{equation}
The product of the eigenvalues of $X$ from the two transverse modes is
\begin{equation}
 \big\{\zeta_1^2+(\sigma\alpha)^2\big\}\big\{\zeta_2^2+(\sigma\alpha)^2\big\}
 ~=~\big\{\Delta_0-\tfrac{2i}f(\sigma\alpha)+(\sigma\alpha)^2\big\}
    \big\{\Delta_0+\tfrac{2i}f(\sigma\alpha)+(\sigma\alpha)^2\big\}.
\end{equation}

Up to now it was assumed that $|m|\le j-1$ and therefore $j\ge1$.
When $m=j$ and $j\ge1/2$, one is interested in the eigenmodes for which
$A$ has no component proportional to $e^-$, so that now $X$
becomes a $3\times 3$ matrix. Its three eigenvalues are
\begin{equation}
 0,\quad\Delta_0+(\sigma\alpha)^2,\quad
 \Big(\tfrac2f+\tfrac{2j}{\tilde\ell}\Big)^2+(\sigma\alpha)^2.
\end{equation}
Similarly, when $m=-j$ and $j\ge1/2$ the matrix $X$ becomes $3\times 3$,
and obtains the same three eigenvalues as above.
When $m=\pm(j+1)$ and $j\ge1/2$ one finds that $X$ becomes one-dimensional,
\begin{equation}
 X~=~\Big(\tfrac{2j+2}{\tilde\ell}\Big)^2+(\sigma\alpha)^2.
\end{equation}
Finally, for $j=0$ the four eigenmodes and eigenvalues of $\Delta_B$
are given by
\begin{eqnarray}
 (A=e^\pm,~\varphi=0) && \Delta_B=\tfrac4{\tilde\ell^2}+(\sigma\alpha)^2,
 \nonumber \\
 (A=e^3,~\varphi=0) && \Delta_B=\tfrac4{f^2}+(\sigma\alpha)^2,
 \nonumber \\
 (A=0,~\varphi=1) && \Delta_B=0.
\end{eqnarray}

Almost all the zero eigenvalues of $\Delta_B$ correspond to gauge
symmetry, so they should be excluded from the physical determinant. The
only exception is the constant mode of $\varphi$, which correspond to
shifting the saddle point and therefore should also be excluded. To
evaluate the partition function correctly, one also has to take proper
account of Faddeev-Popov determinant. Instead of introducing ghost
fields and modifying the supersymmetry by BRST transformation, we choose
to take a quicker route which takes advantage of saddle point approximation.

Now that we worked out the spectrum and eigenmode decomposition of
the operator $\Delta_B$, one can regard the path integral (before gauge
fixing) as an integral over the mode variables $x_i$ and $\tilde x_j$
corresponding to zero and nonzero eigenmodes of $\Delta_B$,
\begin{equation}
 DAD\varphi~=~
  \prod_{(\Delta_B=0)}\!\!dx_i~\times
  \prod_{(\Delta_B\ne0)}\!\!d\tilde x_j.
\end{equation}
We assume the mode variables are normalized to satisfy
\begin{equation}
\frac12\int\text{Tr}\big(
 A\wedge\ast A+\varphi\wedge\ast\varphi\big)~=~
 \pi\sum_ix_i^2
+\pi\sum_j\tilde x_j^2,
\label{normmode}
\end{equation}
so that the following equality can be reproduced using variables
$x_i,\tilde x_j$,
\begin{equation}
 \int DAD\varphi\exp\left(-\frac12\int\text{Tr}\big(
 A\wedge\ast A+\varphi\wedge\ast\varphi\big)\right)~=~1.
\label{mes}
\end{equation}
Excluding the zero eigenvalues from the determinant of $\Delta_B$
corresponds to the insertion of $\prod\delta(x_i)$. But the correct gauge
fixing is given by $\prod\delta(\omega_i)={\cal J}\prod\delta(x_i)$, where
$\omega_i$ are the mode variables of gauge transformation which acts on
the fields $A$ and $\varphi$ as
\begin{equation}
 \delta_\omega A ~=~ d\omega,
 \quad
 \delta_\omega\varphi ~=~i[\omega,\sigma]
\label{domega}
\end{equation}
on the saddle point labelled by $\sigma$. The variables $\omega_i$ also
satisfy the normalization condition similar to (\ref{normmode}). The
Jacobian ${\cal J}$ for the change of variables is called Faddeev-Popov
determinant. Noticing $\prod dx_i={\cal J}D\omega$, one can determine
${\cal J}$ by inserting (\ref{domega}) into (\ref{mes}),
\begin{equation}
 1~=~ {\cal J}\cdot
 \int D'\omega\exp\left(-\frac12\int d^3\xi\sqrt{g}\;\text{Tr}\big(
 \partial_\mu\omega\partial^\mu\omega-[\sigma,\omega]^2\big)\right).
\end{equation}
Note that the constant mode should be excluded from the measure
$D'\omega$ as explained in the previous paragraph. This path integral
can be easily worked out using spherical harmonics. The mode $\omega\sim
Y_{j,m}E_\alpha$ is an eigenmode of the Laplace operator with the eigenvalue
\begin{equation}
 \Delta_0(j,m)+(\sigma\alpha)^2.
\end{equation}
This precisely cancels with the contribution of longitudinal vector
eigenmode to the determinant. The path integral over bosons
$A_{\pm\alpha},\varphi_{\pm\alpha}$ modulo gauge equivalence finally becomes
\begin{eqnarray}
&&
 \Big(\tfrac4{f^2}+(\sigma\alpha)^2\Big)^{-1}
 \prod_{j\ge1/2}
 \Big[\big(\tfrac2f+\tfrac{2j}{\tilde\ell}\big)^2+(\sigma\alpha)^2\Big]^{-2(2j+1)}
 \prod_{j\ge0}
 \Big[\big(\tfrac{2j+2}{\tilde\ell}\big)^2+(\sigma\alpha)^2\Big]^{-2(2j+1)}
 \nonumber \\ &&\times
 \prod_{j\ge1}\prod_{m=1-j}^{j-1}
 \Big(\Delta_0(j,m)-\tfrac{2i}f\sigma\alpha+(\sigma\alpha)^2\Big)^{-2j-1}
 \Big(\Delta_0(j,m)+\tfrac{2i}f\sigma\alpha+(\sigma\alpha)^2\Big)^{-2j-1}.
\end{eqnarray}
where $j$ runs over half-integers.

Combining the bosonic and fermionic contributions together, we obtain
the one-loop determinant of vector multiplet.
\begin{eqnarray}
 \prod_{\alpha\in\Delta_+}\prod_{n>0}
 \Big[\frac{n^2}{\tilde\ell^2}+(\sigma\alpha)^2\Big]^2
 ~\sim~  \prod_{\alpha\in\Delta_+}
 \Big(\frac{\sinh(\pi\tilde\ell\sigma\alpha)}{\pi\tilde\ell\sigma\alpha}\Big)^2.
\end{eqnarray}
Here the product runs over all positive roots $\alpha$. The factors in the
denominator cancel against the Vandermonde determinant which arises when
the integration over the Lie algebra is reduced to that on
Cartan subalgebra. Again, the result is essentially the same as for
round $S^3$ \cite{KWY1} except now $\tilde\ell$ is playing the role of the
radius of round $S^3$.

\vskip2mm

Thus, after all these tedious computations, we found a rather
disappointing result that the familiar squashing of $S^3$ with
$SU(2)_{\mathscr{L}}\times U(1)_{\mathscr{R}^3}$ gives nothing new.
We notice here that there is a simple reason for this. Since our
squashed $S^3$ has an unbroken $SU(2)$ symmetry, the eigenmodes of
Laplace or Dirac operators naturally form multiplets with the same
eigenvalues. The exponent $n$ in the formula (\ref{detmat1}) simply
reflects the fact that the multiplicity becomes larger as the angular
momentum $j$ gets larger. This is tied to the degeneration of zeroes and
poles of $s_b(x)$ for $b=1$. Therefore, in order to find generalization
to $b\neq1$, we need to look for a less symmetric squashing of $S^3$.

\section{Partition function on Squashed $\bf S^3$ (Less Familiar One)}
\label{sec:partition2}

In this section we study the partition function on less familiar
version of squashed $S^3$ which preserves only
$U(1)_{\mathscr{L}^3}\times U(1)_{\mathscr{R}^3}$ symmetry.
The vielbein, spin connection, Killing spinors and the background
$U(1)$ gauge field $V$ are summarized in section \ref{sec:three-sphere}.
In the following discussion we regard the Killing spinors $\epsilon,
\bar\epsilon$ to be Grassmann-even. For convenience, we also renormalize
them by constants so that they satisfy
\begin{equation}
 \bar\epsilon\epsilon=1,\quad
 v^\mu v_\mu=
 \bar\epsilon\gamma^\mu\epsilon\cdot
 \bar\epsilon\gamma_\mu\epsilon=1.
\end{equation}
More explicitly, we choose
\begin{equation}
 \epsilon=\frac1{\sqrt2}\left(\begin{array}{r}
  -e^{\frac i2(\chi-\varphi+\theta)}\\
   e^{\frac i2(\chi-\varphi-\theta)}\end{array} \right),\quad
 \bar\epsilon=\frac1{\sqrt2}\left(\begin{array}{r}
   e^{\frac i2(-\chi+\varphi+\theta)}\\
   e^{\frac i2(-\chi+\varphi-\theta)}\end{array} \right),
\end{equation}
so that
\begin{eqnarray}
 \bar\epsilon\gamma^a\epsilon &=&
 (-\cos\theta,\sin\theta,0), \nonumber \\
 \epsilon\gamma^a\epsilon &=&
 (i\sin\theta,i\cos\theta,+1)e^{i(\chi-\varphi)}, \nonumber \\
 \bar\epsilon\gamma^a\bar\epsilon &=&
 (i\sin\theta,i\cos\theta,-1)e^{-i(\chi-\varphi)}.
\end{eqnarray}
Note also the equalities
\begin{equation}
 v_\mu\bar\epsilon\gamma^\mu~=~\bar\epsilon,\quad
 \gamma^\mu\epsilon v_\mu=\epsilon,
\end{equation}
which will be frequently used in what follows.

Since the squashed $S^3$ of our interest here has a reduced symmetry and
the metric cannot be written in terms of LI or RI forms, it will be a
tedious task to work out all the eigenmodes of the bosonic and fermionic
kinetic operators. In fact, the precise form of most of the eigenmodes
and eivenvalues is irrelevant since, as we have seen in the previous
section, they give cancelling contributions to the determinant due to
the pairing of bosonic and fermionic modes by supersymmetry. Therefore,
to compute the one-loop determinant, it will be useful to understand how
this cancellation happens.

\paragraph{Matter multiplets.}

We study the matter fields first, focusing on the case with
a single chiral multiplet coupled to an abelian vector multiplet.
What we need is the spectrum of the kinetic operators $\Delta_\phi$ and
$\Delta_\psi$ for bosons and fermions. In this section we define them
from the regulator Lagrangian
\begin{eqnarray}
 {\cal L}_\text{reg}&=& \delta_{\bar\epsilon}\delta_\epsilon
 \Big(\bar\psi\psi-2i\bar\phi\sigma\phi\Big)
 \nonumber \\ &=&
  D_\mu\bar\phi D^\mu\phi
 +\tfrac{2i(q-1)}f v^\mu D_\mu\bar\phi\phi
 +\bar\phi\sigma^2\phi
 +i\bar\phi\big(\tfrac{\sigma}f+D\big)\phi
 +\tfrac{2q^2-3q}{2f^2}\bar\phi\phi
 +\tfrac q4R\bar\phi\phi
 \nonumber \\ &&
 -i\bar\psi\gamma^\mu D_\mu\psi
 +i\bar\psi\sigma\psi
 -\tfrac{1}{2f}\bar\psi\psi
 +\tfrac{q-1}f\bar\psi\gamma^\mu v_\mu\psi
 +i\bar\psi\lambda\phi
 -i\bar\phi\bar\lambda\psi
 +\bar FF.
\label{Lreg}
\end{eqnarray}
This leads to the kinetic operators
\begin{eqnarray}
 \Delta_\phi &=& -D_\mu D^\mu-\tfrac{2i(q-1)}fv^\mu D_\mu+\sigma^2
 +\tfrac{2q^2-3q}{2f^2}+\tfrac{qR}4,
\nonumber \\
 \Delta_\psi &=& -i\gamma^\mu D_\mu+i\sigma-\tfrac{1}{2f}
 +\tfrac{q-1}f\gamma^\mu v_\mu,
\end{eqnarray}
acting on scalars and spinors of R-charges $-q, 1-q$ respectively.
Note that in deriving $\Delta_\phi$ we used
\begin{equation}
D_\mu v^\mu=0,\quad
 v^\mu \partial_\mu f=0,
\end{equation}
which can be shown using Killing spinor equation.

Now, it is a tedious but straightforward computation to show the following.
First, if $\Psi$ be a spinor eigenmode for $\Delta_\psi=M$, then
$\bar\epsilon\Psi$ is a scalar eigenmode for $\Delta_\phi=M(M-2i\sigma)$.
Second, let $\Phi$ be a scalar eigenmode for $\Delta_\phi=M(M-2i\sigma)$.
Then if we define a pair of spinor wave functions as
\begin{equation}
 \Psi_1 = \epsilon\Phi,\quad
 \Psi_2 = i\gamma^\mu\epsilon D_\mu\Phi
 +i\epsilon\sigma\Phi-\tfrac qf\epsilon\Phi,
\label{btof}
\end{equation}
then $\Delta_\psi$ acts on them as follows.
\begin{equation}
 \left(\begin{array}c \Delta_\psi\Psi_1 \\ \Delta_\psi\Psi_2 \end{array}\right)
~=~\left(\begin{array}{cr}
 2i\sigma & -1 \\ -M(M-2i\sigma) & 0 \end{array}\right)
\left(\begin{array}c \Psi_1 \\ \Psi_2 \end{array}\right).
\end{equation}
The $2\times 2$ matrix on the right hand side has eigenvalues $M$
and $2i\sigma-M$. We thus found a pairing between a scalar eigenmode
with $\Delta_\phi=M(M-2i\sigma)$ and two spinor eigenmodes with
$\Delta_\psi=M,\;2i\sigma-M$. Any modes which take part in this pairing
can be neglected when computing the one-loop determinant.
Note that part of this pairing, namely $\Phi=\bar\epsilon\Psi$ and the
construction (\ref{btof}) of $\Psi_2$ from $\Phi$, could be guessed from
the supersymmetry transformation rules (\ref{dncm}).

Had we chosen the Lagrangian (\ref{Lncm}) for the regulator, we would
find that the relation between scalar and spinor eigenmodes contains
$f$ and therefore becomes coordinate-dependent. This can be traced
to the third term in the right hand side of the first equation in
(\ref{Ltot}). Since one may use arbitrary regulator Lagrangian
as long as it regulates the integral over all the modes transverse to
the saddle point locus, we may choose it so that the relation between
bosonic and fermionic eigenmodes becomes as simple as possible. This is
why we chose the Lagrangian (\ref{Lreg}).

Nontrivial contributions to one-loop determinant come from the modes
which do not fall into the multiplet structure explained in the last
paragraph. There are two types of such modes.

The first is unpaired spinor eigenmodes $\Psi$, which vanish
when contracted with $\bar\epsilon$ and therefore do not have scalar
partners. Such modes can be expressed as $\Psi=\bar\epsilon F$, where
$F$ is a scalar with R-charge $2-q$. The eigenmode equation
$\Delta_\psi\Psi=M\Psi$ can be rewritten into
\begin{eqnarray}
 \big(M+\tfrac{q-2}f-i\sigma\big)F &=&
 -i\cos\theta D_1F+i\sin\theta D_2F,
 \nonumber \\
 D_3F &=& +i\sin\theta D_1F+i\cos\theta D_2F,
\label{detF}
\end{eqnarray}
where we have used the explicit form of $\bar\epsilon$, and
\begin{eqnarray}
 D_1F &=& \frac1{\ell\cos\theta}
 \Big\{\partial_\varphi-\tfrac i2(q-2)(1-\tfrac\ell f)\Big\}F,
 \nonumber \\
 D_2F &=& \frac1{\tilde\ell\cos\theta}
 \Big\{\partial_\chi+\tfrac i2(q-2)(1-\tfrac{\tilde\ell}f)\Big\}F,
 \nonumber \\
 D_3F &=& \frac1f\partial_\theta F.
\end{eqnarray}
If we assume $F\sim e^{im\varphi-in\chi}$, then the first equation in
(\ref{detF}) determines the eigenvalue $M$,
\begin{equation}
 M~=~ i\sigma+\frac m\ell +\frac n{\tilde\ell}
 -\frac{q-2}2\Big(\frac1\ell+\frac1{\tilde\ell}\Big),
\end{equation}
while the second equation determines the $\theta$-dependence of $F$.
\begin{equation}
 \frac1f\partial_\theta F ~=~
 -\frac{\sin\theta}{\ell\cos\theta}
  \Big\{m-\tfrac{q-2}2(1-\tfrac\ell f)\Big\}F
 +\frac{\cos\theta}{\tilde\ell\sin\theta}
  \Big\{n-\tfrac{q-2}2(1-\tfrac{\tilde\ell}f)\Big\}F.
\label{detF2}
\end{equation}
The precise form of the solutions to this differential equation is not
important, but we need to know what values of $(m,n)$ leads to normalizable
eigenmodes.
We recall here that the squared norm of $F$ is defined by an integral
over $\theta\in[0,\pi/2]$ with a measure which is proportional to
$d\theta f(\theta)\sin\theta\cos\theta$. The solution to (\ref{detF2})
may develop singularities at the two ends of the integration domain.
The behavior of $F$ there is easily found to be
\begin{equation}
 F \sim \cos^m\theta\sin^n\theta.
 \quad(\theta\sim 0~~\text{or}~~\pi/2)
\end{equation}
Therefore the normalizability requires $m,n$ to be nonnegative.

The second is the missing spinor eigenmodes. This is the case where
the map (\ref{btof}) does not give two independent spinor eigenmodes
from one scalar eigenmode. Namely $\Psi_1,\Psi_2$ are proportional
to each other. Let us put $\Psi_2=M\Psi_1$, i.e.
\begin{equation}
 i\gamma^\mu\epsilon D_\mu\Phi+i\epsilon\sigma\Phi-\tfrac qf\epsilon\Phi
 ~=~ M\epsilon\Phi.
\label{detPhi}
\end{equation}
Then one can show that $\Psi_2=M\Psi_1$ is a spinor eigenmode for
$\Delta_\psi=2i\sigma-M$, and moreover $\Phi$ has the eigenvalue
$\Delta_\phi=M(M-2i\sigma)$. That is to say, a spinor eigenmode for
$\Delta_\psi=M$ is missing.
To work out the spectrum of missing eigenvalues, we rewrite
(\ref{detPhi}) into the following form,
\begin{eqnarray}
 \big(M+\tfrac qf-i\sigma\big)\Phi &=&
 -i\cos\theta D_1\Phi+i\sin\theta D_2\Phi,
 \nonumber \\
 D_3\Phi &=& -i\sin\theta D_1\Phi-i\cos\theta D_2\Phi.
\end{eqnarray}
The rest of the computation is the same as for unpaired spinor eigenmodes.
The missing spinor eigenvalues are thus given by
\begin{equation}
 M~=~ i\sigma-\frac m\ell -\frac n{\tilde\ell}
 -\frac q2\Big(\frac1\ell+\frac1{\tilde\ell}\Big),\quad
 m,n\ge0.
\end{equation}

The one-loop determinant is the product of all the unpaired spinor
eigenvalues divided by the product of all the missing spinor
eigenvalues. Ignoring the sign factors we find
\begin{equation}
 \frac{\text{det}\Delta_\psi}
      {\text{det}\Delta_\phi}
  ~=~ \prod_{m,n\ge0}
 \frac{\frac m\ell+\frac n{\tilde\ell}+i\sigma
       -\frac{q-2}2(\frac1\ell+\frac1{\tilde\ell})}
      {\frac m\ell+\frac n{\tilde\ell}-i\sigma
       +\frac q2(\frac1\ell+\frac1{\tilde\ell})}.
\end{equation}
Introducing
\begin{equation}
 b\equiv(\tilde\ell/\ell)^{\frac12},\quad
 Q\equiv b+b^{-1},\quad
 \hat\sigma\equiv (\ell\tilde\ell)^{\frac12}\sigma,
\label{Lipar}
\end{equation}
one can rewrite this using the double sine function $s_b(x)$
given in the Introduction,
\begin{equation}
 \frac{\text{det}\Delta_\psi}
      {\text{det}\Delta_\phi}
  ~=~ \prod_{m,n\ge0}
 \frac{mb+nb^{-1}+\frac Q2+i\hat\sigma+\frac Q2(1-q)}
      {mb+nb^{-1}+\frac Q2-i\hat\sigma-\frac Q2(1-q)}
 ~=~ s_b\big(\tfrac{iQ}2(1-q)-\hat\sigma\big).
\label{nicedet1}
\end{equation}
Now $b$ is determined from the shape of the background squashed sphere
and it can take any values, which is precisely what we have been seeking for!

A comment on ${\cal N}=4$ extended supersymmetry is in order. Recall a
${\cal N}=4$ hypermultiplet is a pair of ${\cal N}=2$ chiral multiplets
$(q,\tilde q)$ with R-charge $1/2$ and opposite gauge charges. Its
one-loop determinant becomes very simple on round sphere ($b=1$),
\begin{equation}
 s_b(\tfrac{iQ}4-\hat\sigma)s_b(\tfrac{iQ}4+\hat\sigma)\Big|_{b=1}
~=~
 s_1(\tfrac i2-\hat\sigma)s_1(\tfrac i2+\hat\sigma)
~=~\frac1{2\cosh\pi\hat\sigma}.
\end{equation}
This follows from an identity of double sine function,
\begin{equation}
 s_b(\tfrac{ib}2-x)s_b(\tfrac{ib}2+x)~=~\frac{1}{2\cosh\pi bx}.
\end{equation}
However, this simplification does not happen on squashed sphere, unless
one turns on a mass deformation by gauging the global symmetry under
which $q,\tilde q$ have the same charge. Therefore, matrix models on
the squashed $S^3$ will be much more complicated than on round $S^3$.

\paragraph{Vector multiplets.}

Next we study the vector multiplets. As in the previous section we denote
by $\varphi$ the fluctuation mode of the scalar field away from its
classical value $\sigma$, and use the original Lagrangian for the
regulator which becomes (\ref{Llin}) after Gaussian approximation.
We also decompose all the adjoint fields into Cartan-Weyl basis.
The modes proportional to the root $\alpha$ get a mass $\sigma\alpha$,
and we denote them with suffix $\alpha$. We thus need to find out what
kind of multiplet structure is formed by the bosonic eigenmode of
$(A_\alpha,\varphi_\alpha)$ and the fermionic eigenmodes of
$\lambda_\alpha$.

We saw in the previous section that, for generic quantum number of
$SU(2)_\mathscr{L}\times SU(2)_{\mathcal{R}}$, the four bosonic
eigenmodes for $(A_\alpha,\varphi_\alpha)$ split into two transverse
modes with $dA_\alpha=0$ and two longitudinal modes with
$A_\alpha \sim d\varphi_\alpha$. After fixing a gauge and combining
with the volume of the gauge group, the longitudinal modes were shown
to yield no net contribution to the one-loop determinant. This property
can be shown to be independent of the 3D metric. Our problem is therefore how
the remaining two transverse modes are paired with the spinor eigenmodes.

The eigenmodes are generically paired in the following manner.
Let $\Lambda$ and $\mathscr{A}$ be spinor and transverse vector
eigenmodes for the eigenvalue $M$ satisfying
\begin{eqnarray}
 M\Lambda &=& \big(i\gamma^\mu D_\mu+i\sigma\alpha-\tfrac1{2f}\big)\Lambda,
\label{MLambda}
 \\
 M\mathscr{A} &=&i\sigma\alpha\mathscr{A}-\ast d\mathscr{A}.
\label{MA}
\end{eqnarray}
Then, up to constant multiplication, they are mapped to each other by
\begin{equation}
 \mathscr{A}\equiv
 d(\bar\epsilon\Lambda)+(iM+\sigma\alpha)
 \bar\epsilon\gamma_\mu\Lambda d\xi^\mu,
 \quad
 \Lambda\equiv \gamma^\mu\epsilon \mathscr{A}_\mu.
\label{ALMap}
\end{equation}
This relation could again be guessed from supersymmetry transformation
rules (\ref{trvec}). First, inserting a transverse vector eigenmode
${\cal A}$ satisfying (\ref{MA}) and a constant $\sigma$ into the right
hand side of $\delta\lambda$ in (\ref{trvec}) gives
\begin{equation}
 \delta\lambda~=~\tfrac12\gamma^{\mu\nu}\epsilon
 (\partial_\mu{\cal A}_\nu-\partial_\nu{\cal A}_\mu)
 -(\sigma\alpha)\gamma^\mu\epsilon{\cal A}_\mu,
\end{equation}
which is proportional to $\Lambda=\gamma^\mu\epsilon{\cal A}_\mu$.
On the other hand, the naive map from spinor to vector eigenmodes is
${\cal A}\simeq \bar\epsilon\gamma_\mu\Lambda d\xi^\mu$, but it can be
shown to satisfy
\begin{equation}
 -i\ast d(\bar\epsilon\gamma_\mu\Lambda d\xi^\mu)
 ~=~
 d(\bar\epsilon\Lambda)
 +(iM+\sigma\alpha)\bar\epsilon\gamma_\mu\Lambda d\xi^\mu,
\end{equation}
namely it fails to satisfy (\ref{MA}) due to a nonzero divergence term.
The right hand side of this equality is by construction divergenceless,
and can be used to define the map from $\Lambda$ to ${\cal A}$.

The modes which have a superpartner under the relation (\ref{ALMap}) are
irrelevant in computing the one-loop determinant. Unpaired spinor eigenmodes
satisfy (\ref{MLambda}) and are annihilated by (\ref{ALMap}), and
contribute the eigenvalue $M$ to the enumerator of the determinant.
Missing spinor eigenmodes correspond to $\mathscr A$ satisfying
(\ref{MA}) and are annihilated by (\ref{ALMap}), and contribute $M$
to the denominator.

We solve the equation for unpaired spinor eigenmodes by putting
\begin{equation}
 \Lambda ~=~ \epsilon\Phi_0+\bar\epsilon\Phi_2,
\end{equation}
where $\Phi_0$ and $\Phi_2$ are scalars with R-charges $0$ and $2$.
The equations for $\Phi_0$ and $\Phi_2$ at first look overdetermined
but turn out to have solutions.
If we put the ansatz
\begin{equation}
 \Phi_0=\varphi_0(\theta)e^{im\varphi-in\chi},\quad
 \Phi_2=\varphi_2(\theta)e^{i(m-1)\varphi-i(n-1)\chi},
\end{equation}
then the solution is determined by
\begin{equation}
 M=\frac m\ell+\frac n{\tilde\ell}+i\sigma\alpha,\quad
 \partial_\theta\varphi_0=
 if\cdot\Big(\frac m\ell+\frac n{\tilde\ell}\Big)\varphi_2,
\nonumber \\
\end{equation}
\begin{equation}
 \Big(\frac1f\partial_\theta+\frac{m\sin\theta}{\ell\cos\theta}
 -\frac{n\cos\theta}{\tilde\ell\sin\theta}\Big)\varphi_0~=~0.
\end{equation}
Normalizability of the solution requiers $m,n$ to be nonnegative.

Also, note that for $m=n=0$ or $M=i\sigma\alpha$ the second
term in the first equation of (\ref{ALMap}) vanishes, so that it does
not give a map from spinor eigenmode to transverse vector eigenmode.
By a direct check one finds for $m=n=0$ there is no normalizable
unpaired spinor eigenmode.

To find the transverse vector eigenmodes with missing spinor partners,
we begin by solving $\mathscr{A}_a\gamma^a\epsilon=0$ by
\begin{equation}
 \mathscr A_1=iY\sin\theta,\quad
 \mathscr A_2=iY\cos\theta,\quad
 \mathscr A_3=Y.
\end{equation}
It is straightforward to solve the eigenmode equation (\ref{MA}) in
components. Putting the ansatz $Y=y(\theta)e^{im\varphi-in\chi}$
one finds
\begin{eqnarray}
 M &=& \frac m\ell+\frac n{\tilde\ell}+i\sigma\alpha,\nonumber \\
 0 &=& \Big(\frac1f\partial_\theta
 -\frac{\sin\theta}{\cos\theta}\big(\frac m\ell+\frac1f\big)
 +\frac{\cos\theta}{\sin\theta}\big(\frac n{\tilde\ell}+\frac1f\big)\Big)y.
\end{eqnarray}
Normalizability of the solution requires $m,n\le-1$.

Now we combine all the contributions to compute the one-loop
determinant.
\begin{eqnarray}
\lefteqn{
 \prod_{\alpha\in\Delta}
 \Bigg[
 \frac1{i\sigma\alpha}
 \prod_{m,n\ge0}
 \frac{\frac m\ell+\frac n{\tilde\ell}+i\sigma\alpha}
      {\frac {-m-1}\ell+\frac {-n-1}{\tilde\ell}+i\sigma\alpha}
 \Bigg]
} \nonumber \\ &=&
 \prod_{\alpha\in\Delta_+}
 \prod_{n>0}
 \Big(\tfrac{n^2}{\ell^2}+(\sigma\alpha)^2\Big)
 \Big(\tfrac{n^2}{\tilde\ell^2}+(\sigma\alpha)^2\Big)
 ~=~ \prod_{\alpha\in\Delta_+}
 \frac{\sinh(\pi b\hat\sigma\alpha)\sinh(\pi b^{-1}\hat\sigma\alpha)}
      {(\pi\hat\sigma\alpha)^2}.
\label{nicedet2}
\end{eqnarray}
Here the product is over all the positive roots $\alpha$, and we used the
parameters of (\ref{Lipar}). The factors in the denominator in the right
hand side cancel against the Vandermonde determinant when reducing the
integral from Lie algebra to its Cartan subalgebra.

\vskip2mm

We thus found that, by putting the 3D theory on the $U(1)\times U(1)$
symmetric squashed $S^3$ (or the hyper-ellipsoid in $\mathbb{R}^4$ with
the four axis-length parameters $\ell,\ell,\tilde\ell,\tilde\ell$), the
partition function is expressed in terms of the double sine function with
$b=(\tilde\ell/\ell)^{\frac12}$, namely the formula (\ref{nicedet1}).
Also, the integration measure over the Coulomb branch (\ref{nicedet2})
is given by products of pairs of $\sinh$ functions. These are both
identified with the building blocks of structure constants in Liouville
or Toda CFTs.

\section{Concluding Remarks}

As was studied in \cite{DGG,HLP}, certain 3D supersymmetric gauge
theories arise on domain walls in 4D ${\cal N}=2$ gauge theories.
When the fields in the 4D theories on the two sides are connected on the
wall via S-duality, the AGT relation relates the 3D partition
functions on the wall with the kernels of the corresponding S-duality
transformation acting on conformal blocks. By replacing the round metric
of the wall $S^3$ by a squashed one, one obtains the kernels for general $b$.
Also, the AGT relation was interpreted in \cite{NW} as the equivalence
between the space of supersymmetric ground states of 4D ${\cal N}=2$ gauge
theories on $\mathbb R\times S^3$ (with omega deformation) and the space
of conformal blocks. Our results should give a generalization of the AGT
relation to $b\ne 1$ along this line, too.

We conclude with one immediate conjecture: there should be a similar
squashing of $S^4$, with some background gauge field turned on, which
gives the generalization of AGT relation to $b\ne 1$. The metric
and gauge field should be such that, when we view $S^4$ as a bundle of
$S^3$ fibred over a line segment, they take precisely the forms given in
this paper when restricted to each fiber.


\vskip1cm

\begin{thebibliography}{999}

\bibitem{KWY1}
  A. Kapustin, B. Willett and I. Yaakov,
  ``{\it Exact Results for Wilson Loops in Superconformal
  Chern-Simons Theories with Matter},''
  JHEP {\bf 1003}, 089 (2010)
  [arXiv:0909.4559 [hep-th]].

\bibitem{KWY2}
  A. Kapustin, B. Willett and I. Yaakov,
  ``{\it Nonperturbative Tests of Three-Dimensional Dualities},''
  arXiv:1003.5694 [hep-th].

\bibitem{Jafferis}
  D.L.~Jafferis,
  ``{\it The Exact Superconformal R-Symmetry Extremizes $Z$},''
  arXiv:1012.3210 [hep-th].

\bibitem{HHL}
  N. Hama, K. Hosomichi and S. Lee,
  ``{\it Notes on SUSY Gauge Theories on Three-Sphere},''
  arXiv:1012.3512 [hep-th].

\bibitem{Suy}
  T. Suyama,
  ``{\it On Large N Solution of ABJM Theory},''
  Nucl. Phys. B {\bf 834}, 50 (2010)
  [arXiv:0912.1084 [hep-th]].

\bibitem{RS}
  S.J. Rey and T. Suyama,
  ``{\it Exact Results and Holography of Wilson Loops in $N=2$
     Superconformal (Quiver) Gauge Theories},''
  arXiv:1001.0016 [hep-th].

\bibitem{MPP}
  M. Marino, S. Pasquetti and P. Putrov,
  ``{\it Large N duality beyond the genus expansion},''
  JHEP {\bf 1007}, 074 (2010)
  [arXiv:0911.4692 [hep-th]].

\bibitem{DT}
  N. Drukker and D. Trancanelli,
  ``{\it A supermatrix model for $N=6$ super Chern-Simons-matter theory},''
  JHEP {\bf 1002}, 058 (2010)
  [arXiv:0912.3006 [hep-th]].

\bibitem{MP}
  M. Marino and P. Putrov,
  ``{\it Exact Results in ABJM Theory from Topological Strings},''
  JHEP {\bf 1006}, 011 (2010)
  [arXiv:0912.3074 [hep-th]].

\bibitem{DMP}
  N. Drukker, M. Marino and P. Putrov,
  ``{\it From weak to strong coupling in ABJM theory},''
  arXiv:1007.3837 [hep-th].

\bibitem{Seok}
  S. Kim,
  ``{\it The complete superconformal index for $N=6$ Chern-Simons theory},''
  Nucl. Phys. B {\bf 821}, 241 (2009)
  [arXiv:0903.4172 [hep-th]].

\bibitem{IY}
  Y. Imamura and S. Yokoyama,
  ``{\it Index for three dimensional superconformal field theories
         with general R-charge assignments},''
  arXiv:1101.0557 [hep-th].

\bibitem{IYY}
  Y. Imamura, D. Yokoyama and S. Yokoyama,
  ``{\it Superconformal index for large N quiver Chern-Simons theories},''
  arXiv:1102.0621 [hep-th].

\bibitem{CGKP}
  S. Cheon, D. Gang, S. Kim and J. Park,
  ``{\it Refined test of AdS4/CFT3 correspondence for $N=2,3$ theories},''
  arXiv:1102.4273 [hep-th].


\bibitem{Pe}
  V. Pestun,
  ``{\it Localization of Gauge Theory on a Four-sphere and Supersymmetric
  Wilson Loops},''
  arXiv:0712.2824 [hep-th].

\bibitem{AGT}
  L.F. Alday, D. Gaiotto and Y. Tachikawa,
  ``{\it Liouville Correlation Functions from Four-dimensional Gauge Theories},''
  Lett. Math. Phys. {\bf 91}, 167 (2010)
  [arXiv:0906.3219 [hep-th]].

\bibitem{Wy}
  N. Wyllard,
  ``{\it $A_{N-1}$ conformal Toda field theory correlation functions
  from conformal ${\cal N}=2$ SU(N) quiver gauge theories},''
  JHEP {\bf 0911}, 002 (2009)
  [arXiv:0907.2189 [hep-th]].

\bibitem{DGG}
  N. Drukker, D. Gaiotto and J. Gomis,
  ``{\it The Virtue of Defects in 4D Gauge Theories and 2D CFTs},''
  arXiv:1003.1112 [hep-th].

\bibitem{HLP}
  K. Hosomichi, S. Lee and J. Park,
  ``{\it AGT on the S-duality Wall},''
  arXiv:1009.0340 [hep-th].

\bibitem{KLS}
  S. Kharchev, D. Lebedev and M. Semenov-Tian-Shansky,
  ``{\it Unitary Representations of $U_q(\mathfrak{sl}(2,\mathbb{R}))$,
  the Modular Double, and the Multiparticle q-deformed Toda Chains},''
  Commun. Math. Phys. {\bf 225}, 573 (2002)
  [arXiv:hep-th/0102180].

\bibitem{BT}
  A.G. Bytsko and J. Teschner,
  ``{\it Quantization of models with non-compact quantum group symmetry:
  Modular XXZ magnet and lattice sinh-Gordon model},''
  J. Phys. A  {\bf 39}, 12927 (2006)
  [arXiv:hep-th/0602093].

\bibitem{NW}
  N. Nekrasov and E. Witten,
  ``{\it The Omega Deformation, Branes, Integrability, and Liouville Theory},''
  JHEP {\bf 1009}, 092 (2010)
  [arXiv:1002.0888 [hep-th]].

\end{thebibliography}
\end{document}